\documentstyle[psfig,sprocl]{article}

\def\be{\begin{equation}}
\def\ee{\end{equation}}
\def\bea{\begin{eqnarray}}
\def\eea{\end{eqnarray}}

\begin{document}

\title{A Color Mutation Hadronic Soft Interaction Model\\
-- Eikonal Formalism and Branching Evolution}

\author{\bf Zhen Cao}

\address{Department of Physics, University of Utah, Salt Lake City UT 84112, USA} 


\maketitle\abstracts{ECOMB is established as a hadronic multiparticle 
production generator by 
soft interaction. It incorporates the eikonal
formalism, parton model, color mutation, branching, resonance production 
and decay. A partonic cluster, being color-neutral initially, splits into 
smaller color-neutral clusters successively due to the color
mutation of the quarks. The process stops at hadronic resonance, $q\bar q$ 
pair, formation. The
model contains self-similar dynamics and exhibits scaling behavior in the
factorial moments, e.g. the intermittency.}

\section{Introduction}

The study of multiparticle production in low-$p_T$ processes has been
pursued for over twenty years.  Since they involve soft interactions,
they cannot be treated in perturbative QCD.  In the absence of any
reliable theoretical approach to the problem, many models have been
proposed, most of which are represented in the review volumes
published ten years ago \cite{car}.  Nearly all of those models have
since been shown to be inadequate in light of the data on fluctuations
and intermittency \cite{ddk}.  Indeed, there are very few models
that have the appropriate dynamical content capable of reproducing
the scaling behaviors observed in the experiments.  
 In this paper we propose a model that
incorporates some aspects of non-perturbative QCD and is capable of
generating the features of the intermittency data, which are shown in
the last figure of this paper.  To our knowledge those data have not
been fitted by any model that contains some features of the color
dynamics.  To reproduce those data has become the main motivator
for this work.

Our approach \cite{cao} embraces many time-honored properties of hadronic
collisions.  Since hadrons are extended objects, the eikonal formalism is
at the foundation of our model.  Thus it is not difficult for our model to
possess the virtues of geometrical scaling, approximate KNO scaling and 
unitarity.  In order to build into the model features of
chromodynamics, it is necessary to introduce quarks and gluons into
the eikonal formalism, so the parton model is an essential gateway
into the microscopic domain of color interactions.  Once we enter that
domain, we embark on an unconventional journey of studying color
mutation of the constituents as a dynamical process by which the
colors of the quarks evolve through the emission and absorption of
gluons.  The smallness of
$\alpha_S$ is never assumed, so the evolution is not perturbative.
With all partons taken into consideration globally, we follow the
evolution of the configuration in the color space.  When the
configuration exhibits color neutral subclusters, we allow branching to
take place with a possible contraction  of the cluster size in
accordance to a reasonable rule consistent with confinement
dynamics. Successive branching leads to smaller and smaller clusters,
until they are finally identified as particles and resonances. The decay of
resonances are also taken into account before the final state of an
event is determined. Evidently, the model contains many features of
soft interaction that are familiar and desirable at a qualitative level.
Here we put them on a quantitative basis.

We shall call this model ECOMB \cite{cao}, which stands for eikonal color
mutation branching. Since the partons play a fundamental role in this
model, it is significantly closer to QCD than any eikonal model on soft
hadronic collisions has ever been.

There are a few parameters in the model. They are adjusted to fit  a
large body of experimental data on low-$p_T$ processes for
$\sqrt{s} \stackrel{<}{\sim} 100$ GeV. They include
$\sigma_{el}$,
$\sigma_{inel}$,
$\left<n\right>$, $C_q$, $dn/dy$, $P_n$, and $F_q$ for all $s$ in the
range
$10 < \sqrt{s} <100$ GeV, and all rapidity intervals. $F_q$ are the
normalized factorial moments, whose power-law dependence on the
rapidity bin-size $\delta$ has been referred to as the intermittency
behavior \cite{bp}.  Except for $F_q$, all the other pieces of data are
global in nature; i.e., examined in or averaged over all rapidity space.
They can be fitted by many models.  $F_q$ in small rapidity intervals
exhibit local fluctuations, which are what invalidate most of those
models.

\section{Geometrical Effect on pp Soft Collision}

To focus on processes in
which hard subprocesses are unimportant, we confine our attention to
the energy range $10<\sqrt{s}<100$ GeV, which covers the CERN ISR
energies. 
From parton model point of view, this soft process with a relative smaller 
momentum transfer undergoes a relative longer time evolution of partons  
than hard subprocess. High energy hadronic collision provides a setting up of the 
initial condition for the partonic evolution. A proper description of the 
collision is the basis of any soft interaction model. Two of the well known
experimental results,  
geometrical scaling and approximate Koba-Neilsen-Olesen (KNO)
scaling, provide immediate constrains on the collision model building. 
The eikonal formalism of
hadronic collisions gives us an ideal geometrical framework to fit all the 
constrains. Here is the highlight of it. 

In terms of the eikonal function $\Omega(b)$, the geometrical scaling, i.e.,
$\sigma_{el}$/$\sigma_{tot}$ is roughly constant, can be guaranteed, 
if $\Omega$ depends only on the scaled impact parameter $R=b/b_0(s)$. The 
reduced inelasticity function
\begin{eqnarray} g(R)=1-e^{-2\Omega(R)}
\label{2.8}
\end{eqnarray} 
satisfies the normalization condition $\int^\infty_0 {d R^2 g(R)}=1$ by setting
$\sigma_{inel}=\pi b^2_0(s)$. 
The function $g(R)$ describes the probability of
having an inelastic collision at $R$. For $pp$ collisions the well-determined
form for the eikonal function is \cite{cy}
\begin{eqnarray}
\Omega(R)=-\ell n\left(1-0.71e^{-1.17R^2}\right) \quad,
\label{2.15}
\end{eqnarray}which has been used to give a good description of
$d\sigma/dt$ \cite{cy}. 

The $\mu$th term of the inelasticity function expansion in a power series,
\begin{eqnarray}
g(R)=\sum^\infty_{\mu=1} \pi_{\mu}(R) = \sum^\infty_{\mu=1} 
{\left[2\Omega (R)\right]^{\mu}\over{\mu !}}e^{-2\Omega (R)},
\label{2.14}
\end{eqnarray}
may be regarded as the
$\mu$th-order rescattering contribution, and can be related to the
$\mu$-cut-Pomeron \cite {hp,ani}. The average $\mu$ is estimated as 1.6 \cite{hp},
according to (\ref{2.14}).

To go further in multiparticle production, it is necessary to model the 
dynamics of soft interaction. This paper is devoted to develop such a 
model by incorporating the parton model into the 
formalism as far as we can. 

To do so,
we adopt the eikonal description to
refer to parton number;
let $B^{\mu}_{\nu}$ be
the probability of having $\nu$ partons in the
${\mu}$-cut Pomeron when the two incident hadrons
are separated by a scaled impact parameter $R$. If we use 
${\cal E}_{\nu\rightarrow n} \{\cdots\}$ to denote  the evolution
process that takes the $\nu$ partons to the $n$ hadrons, we may express
symbolically the multiplicity distribution of hadron as
\begin{eqnarray}
{P}_n =  {\cal E}_{\nu\rightarrow n}\{ \int
dR^2\sum^\infty_{\mu=1}\pi_{\mu}(R)B^{\mu}_\nu \}
\quad.
\label{3.5a}
\end{eqnarray}
The form invoked in the bracket essentially provides a parton number 
distribution which involved in the $\mu$-cut Pomeron exchange soft interaction
and 
the description of ${\cal E}_{\nu\rightarrow n}$ is the main task of this
paper.

A way of thinking about $B^{\mu}_{\nu}$ is to consider the
$\mu = 1$ case, for which we have an one-Pomeron exchange diagram
for the elastic scattering amplitude with the Pomeron being cut to
expose the internal quark, antiquark and gluon lines on mass shell.
We use
$B^{\mu}_{\nu}$ to represent the probability having $\nu$ such partons as 
$\mu$ Pomerons in the elastic amplitude are cut.

Since there exists no rigorous derivation of $B^{\mu}_{\nu}$,
we shall assume that it is Poisson distributed around some mean
number $\bar{\nu}$ of partons.  It is reasonable, since it is known
that a cut ladder corresponds to a multiperipheral diagram for a
$\nu$-particle production amplitude, whose rapidity distribution is
uniform, and multiplicity distribution Poissonian.   The mean number
can depend on both $\mu$ and $s$.Being guided by so-called $\log{s}$ 
physics, we adopt a parameterization for
\begin{eqnarray}
\bar{\nu} (\mu, s) = \hat{\nu} (s) \mu^{a(s)} \quad ,
\label{3.2}
\end{eqnarray}
where the parameters depend on the energy by 
$\hat{\nu}(s)=\nu_0+\nu_1{\rm ln} s$ 
and $a(s)=a_0+a_1{\rm ln}s +a_2{\rm ln}^2 s$. 

Once the number of parton, $\nu$, is determined, one can move on the
issue of evolution of the $\nu$ parton cluster. Before going further, 
I like to point out an  important departure from the usual 
application of the parton model: We combine the property of hadrons being spatially
extended objects with the property of hadrons being made up of
smaller constituents. Clearly, for most collisions whose impact
parameters are nonzero, only portions of the partons in the incident
hadrons overlap and interact, so the number of hadrons produced
depends on that overlap. The event-to-event fluctuation of the
particle multiplicity therefore depends strongly on the
impact-parameter fluctuation, and the event averaged parton
distributions, as determined from the structure functions, are of no
use in that respect.

As one of the consequences of the carefully consideration of event-to-event
fluctuation, not only the average charged multiplicity of hadron 
$\left<n_{ch}\right>$, but also the KNO scaling are reproduced.
They are satisfactorily consistent with the data \cite{dat} in terms of the 
 standard moments $C_q(s) = \left<n^q\right> /\left<n\right>^q$ 
over an energy range of $10 < \sqrt{s} < 70$ GeV as shown in Fig\ 1.

\section{Color Mutation}

We now consider the color dynamics of soft interaction. The initial
state is that there are $\nu$ partons distributed in some fashion in a
linear array in rapidity space. This is a consequence of the parton
model in that the partons are in the incident hadrons to begin with,
and the collision rearranges those partons and sets off the evolution
process that takes those partons to the final state, where the produced
hadrons are decoupled. The evolution process involves quarks and
antiquarks emitting and absorbing gluons. Since the process is not
perturbative, there is no analytical method to track the time development
of the process. Thus we shall use Monte Carlo simulation to generate
the configuration at each time step.

There are two spaces in which we must track the motion of the color
charges. One is the two-dimensional color space; the other is the
one-dimensional rapidity space. The latter can be extended to include
the azimuthal angle
$\phi$ and the transverse momentum $\left|\vec{p}_T\right|$, but in
our first attempt here we integrate over those variables and examine
the simpler problem of a 1D system. As for the color space, it is 2D,
and spanned by the ($I_3$,$Y$) axes, as in $SU(3)$ flavor. 

The
configuration in the two spaces are coordinated in the following sense.
Starting from the extreme left end of the rapidity ($\eta$) space, for
which the system that contains no partons is by definition color neutral
and therefore is
represented by a point at the origin of the  color space, we move in
the positive direction in $\eta$ space until we cross a color
charge. At the point to the right of that color charge, the
corresponding point in the color space jumps from the origin
to a position that represents the color of the charge that has
been crossed. Let us denote that jump by a vector. Then each
time we cross a color charge in the
 $\eta$ space, there is a corresponding vector added to the previous
point in the color space. The succession of additions of those vectors,
each one starting from the tip of the previous one, forms a path. Since
the whole parton system is color neutral, by the time we have
moved to the extreme right in the $\eta$ space, the path in the color
space returns to the origin, thus forming a closed loop. Such a loop
may be self-intersecting at various points. We call such a closed path
a configuration of the system in color space.

In order to initiate the time evolution, we  need distribute the $\nu$ 
partons into both of those two spaces to form such an initial configuration. 

\subsubsection*{(1) Initial Distribution in Rapidity Space}

By rapidity we mean space-time rapidity $\eta$, where  
$\eta = {1\over 2}\, \ell n\, {t + z \over t - z}$, because we are dealing with
a variation of spatial separations between partons 
due to the color force during the time evolution.

The usual parton distribution in momentum fraction (at the lowest accessible
$Q^2$) implies roughly a flat $\eta$ distribution with rapid
damping toward zero at large $\eta$. It is modeled as
\begin{eqnarray}
\rho(\eta) = \rho^{(1)}(\eta) + \rho^{(2)}(\eta)\, ,
\label{29.1}
\end{eqnarray}
\begin{eqnarray}
\rho^{(1)}(\eta) = \rho_0&(&{\eta + \eta_0 \over 2 \eta_0 }\,
\theta(\eta+\eta_0)\, \theta(\eta_0 - \eta) +  
\theta(\eta-\eta_0)\, \theta(\eta_c - \eta) +  \nonumber \\
&&{\eta_{max} - \eta \over \eta_{max} - \eta_c }
\theta(\eta-\eta_c)\, \theta(\eta_{max} -\eta))\,,
\label{29.2}
\end{eqnarray}
\begin{eqnarray}
\rho^{(2)}(\eta) = \rho_0&(&{-\eta + \eta_0 \over 2 \eta_0 }\,
\theta(\eta+\eta_0)\, \theta(\eta_0 - \eta) + 
\theta(\eta+\eta_c)\,\theta(-\eta_0-\eta) +   \nonumber  \\
&&{\eta_{max} + \eta \over \eta_{max} - \eta_c }
\theta(\eta+\eta_{max})\, \theta(-\eta_c-\eta))\,.
\label{29.3}
\end{eqnarray}
in which, $\rho^{(1)}(\eta)$ and $\rho^{(2)}(\eta)$ indicate the forward
and backward semi-spheres respectively. Only small portion of partons overlap
 in a limited central region $(-\eta_0,\eta_0)$ after collision follows 
from the known
fact that the correlation between produced particles is short-ranged, because 
the soft interaction range is short in the conventional wisdom. 


In the color space spanned by $I_3$ and $Y$, a quark is represented
by a vector, which has the coordinates of one of the triplets: $(1/2,
1/3), (-1/2, 1/3)$, and $(0,\,-2/3)$. An antiquark is
represented by a vector directed opposite to one of the above.
As a distribution of partons in the $\eta$ space is generated, say, $
\rho^{(2)}(\eta)$ between $-\eta_{\rm max} \le \eta \le \eta_0$,  we assign
to each parton a color vector
consistent with the requirement that the quarks and antiquarks come in
pairs, but their
orderings in the $\eta$ and the color spaces are totally random.
The partons in $\rho^{(1)}(\eta)$ are distributed similarly, but completely
independently, between $-\eta_0 \le \eta \le \eta_{\rm max}$. Due to the 
merge in the central region $(-\eta_0,\eta_0)$, $\nu$ partons are no longer 
divided into two separate color neutral groups. Usually, they mix up into a 
whole neutral system according to the rule to count the color charge as 
described in the beginning of this section. 

\subsubsection*{(3) Color Interaction}

We now consider the evolution of a configuration due to QCD
dynamics. A nonperturbative treatment of $\nu$ simultaneously
interacting color charges is, of course, too difficult to contemplate here.
We reduce the problem by considering pairwise near-neighbor
interaction via the exchange of a gluon in any of the $s$-, $t$-, or
$u$-channel, whichever is applicable. Starting from the extreme
left in the $\eta$ space, we regard the ordered chain of $\nu$
partons as having $\nu-1$ links (with varying link lengths). 
Pairwise near-neighbor interaction means that we
consider the $\nu-1$ links  in $\eta$ space one at a time, according to a
rule to be specified below. After the interactions at all $\nu-1$ links
are considered, the evolution of the whole configuration is
regarded as having taken one time step, and the process is then
repeated.

For every global configuration $\alpha$ of the $\nu$ partons, there is
an associated energy $E_{\alpha}$. 
$E_{\alpha}$ is determined by
\begin{eqnarray}
 E_{\alpha}=\sum^{\nu -1}_{i=1} \left|\vec{C}_i\right|^2 
= \sum^{\nu -1}_{i=1} \left| \sum^i_{j=1}\vec{c}_j\right|^2 \quad ,
\label{4.7}
\end{eqnarray}
where $\vec{c}_j$ is the color vector for $j$th parton, $\vec{C}_i$ is
the color charge the $i$-link seeing on its left side and sums obviously
depends on the particular path in color space. 
 Whenever a local pair of partons interact, the outcome
may or may not affect the global configuration. Our
statistical approach to the determination of the global configuration
$\alpha$, consistent with favoring the lowest energy state, requires that
the probability for configuration $\alpha$ to occur is
\begin{eqnarray} P_{\alpha}=e^{-\beta E_{\alpha}}/Z, \qquad
Z =
\sum^c_{\alpha=1}e^{-\beta E_{\alpha}},
\label{4.3}
\end{eqnarray} where $\beta$ is a free parameter.
Using the Metropolis algorithm, we use $P_{\alpha}$ to determine the
outcome of a local interaction at every link. 

\subsubsection*{(4) Spatial Fluctuations}

In the meanwhile the system undergoing color mutation due to the exchanging of gluon,
the spatial position of quarks also fluctuates due to the same reason. We track
the change in the length of the $i$th link in the $\eta$ space, 
$d_i=\eta_{i+1}-\eta_i$, to quantify the spatial evolution of $i$th link. 
 Whether the link
length contracts or expands depends on the attractiveness or repulsiveness
of the net color forces that act on the two ends of the link.  To determine the
nature of that force is beyond the scope of this treatment.  We shall model
the change in link length by a stochastic approach, consistent with how we
have handled the color mutation part of the dynamics of the complex
system.   We allow $d_i$ to change by $m_i$, i.e., $d_i \rightarrow
d_i+m_i$, where the probability for $m_i$ is specified by a distribution
\begin{eqnarray}
{\cal P}_i(m) = \gamma_1\, \theta(m+d_i)\, \theta(\gamma_2-m)
+[1-\gamma_1(\gamma_2+d_i)]\, \delta_{m,-d_i} \quad,
\label{4.7c}
\end{eqnarray}
which satisfies $\sum_{m=-d_i}^{\gamma_2}{\cal P}_i(m)=1.$
This distribution allows $m$ to be positive (expansion) uniformly up to
$\gamma_2$, and negative (contraction) down to $-d_i$. $\gamma_1$ and
$\gamma_2$ are two free parameters.

\section{Branching and Hadronization}

A fission of the
color-singlet cluster occurs, when it contains two color-singlet
subclusters, since no confining force exists between them. 
Thereafter, the color mutation process is applied to the two
subsystems separately and independently. This is repeated again and
again until all subclusters consist of only $q\bar{q}$ pairs.
At every stage of the evolution process, a cluster shrinks due to overall
spatial contraction. We always keep the center of the cluster invariant
to conserve momentum. When a branching occurs, the two daughter
clusters will have their own respective centers, which will remain
invariant during contraction, until they themselves branch. At the end
when a hadron is formed from a $q\bar{q}$ pair, the hadron
momentum rapidity $y$ will be identified with a value of
$\eta$ taken randomly between the $\eta$ values of the quark and
antiquark. The reason for doing this is that without knowing the mass
and the transverse momentum $p_T$ of the hadron. A rough identification 
of $y$ with $\eta$ is justified for free particles at high energy. 

When branching process terminates, the $q\bar{q}$ pairs 
are identified as hadrons, which may be pions, kaons, or
resonances. Those resonances must be allowed to decay before the
total number and distribution of particles are counted for the final
state of the event. The probabilities of producing various resonances
and stable (in strong interaction) particles have been studied
experimentally in $pp$ collisions in \cite{res}. We use that reference
as a generic guide for the proportions of all particles produced in any
general hadronic collision. We give each hadron a transverse
momentum according to a ``standard'' exponential distribution of transverse
momentum squared. The average transverse momentum is a parameter to be 
fixed actually as 400MeV. If the
$q\bar{q}$ pair forms a resonance, then we assume an isotropic
decay distribution in the rest frame of the resonance. The
azimuthal angle is assigned randomly. After boosting back to
the cm system, the 3-momenta $\vec{p}$ of the decay particles
are then determined. 

\section{Results}
In the previous section we have introduced six parameters:
$\eta_c$,  $\eta_{\rm max}$ and $\eta_0$ in
(\ref{29.2}) and (\ref{29.3}), $\beta$ in (\ref{4.3}),
$\gamma_1$ and $\gamma_2$ in (\ref{4.7c}).  
They are to be varied to fit the data on inclusive distributions 
and on fluctuations of the exclusive distributions.

The parameters  $\eta_c$, $\eta_{\rm max}$ and $\left< k^2_T\right>=400$MeV
are essentially kinematical; they set the boundaries of the phase space in
which the partons are placed.  They do not affect the dynamics of
color mutation and the spatial fluctuation of the clusters.  The
data used to determine them are the rapidity distribution $dn/d\eta$
\cite{thom} and the
transverse momentum distribution.   The energy range of the data in \cite{thom}
is $22 < \sqrt{s} < 63$   GeV, for which hard scattering is negligible.
The values of $\eta_c$ and $\eta_{\rm max}$ depend on $s$. In Fig.\ 2
we show the pseudorapidity distributions to compare with ECOMB as an example.
The agreement is clearly satisfactory. 

The parameters that influence the evolution process of
mutation and branching are $\eta_0$ , $\beta$, $\gamma_1$ and
$\gamma_2$. $\eta_0$ specifies the overlap region of the initial
color-neutral clusters,
$\beta$ pertains to the probability of color mutation, and
$\gamma_1$ and $\gamma_2$ characterize the fluctuations of the link
lengths. The data used to determine them are on factorial moments $F_q$
for $q=1,2,...$ and their variants \cite{ddk}. 
It is here that we can underline the importance of
those data on fluctuations, without which we have no guidance
on how to restrict the detail dynamics of particle production.
Putting that in another way, in the absence of a procedure to
calculate from first principles, any model that fails to fit the
fluctuation data is missing some aspect of the basic dynamics. To
the extent of our awareness, very few models on soft interaction
have been put to the test of confronting those data on the factorial
moments for varying bin sizes.

The parameters that we have varied to give the best fit of the
intermittency data at ${\sqrt s}=22$ GeV are: $\hat \nu(s) = 9.1, a(s) = 0.63,
\eta_{\rm max}=5, \eta_c=3.5, \eta_0=1.9, 
\beta=0.0015, \gamma_1=0.077, \gamma_2=5$. In Fig.\ 3
we show the intermittency results calculated with ECOMB and data 
of \cite{na2} which to our
knowledge have not been reproduced by any model, except ECCO
\cite{hp,ddk}.  
It should be noted that to achieve the fits attained is highly nontrivial. If
any part of the dynamical process in generating the hadrons is altered,
one would not be able to obtain the rising factorial moments, no matter
how many parameters are used.  By working with the many parts of our
model, all of which affect the determination of $F_q$, we have
gained confidence in regarding the dynamics of color mutation  and
branching as having captured the essential properties of soft interaction.

\section{Conclusion}

It is rather satisfying that we have been able to reproduce the
intermittency data in Fig.\ 3.
Scaling behavior of that type
implies self-similarity in the dynamics of particle production.

We have amalgamated many concepts that form various elements of
the conventional wisdom about soft interaction. They include: (a)
hadrons having sizes, (b) eikonalism, (c) parton model, (d)
interaction of quarks via gluons, (e) statistical properties of a
many-body system, (f) spatial contraction and expansion of a color system,
 and (g) resonance production. They are interlaced by intricate connections
described in this paper.

The most important omission in this paper is Bose-Einstein
correlation, without which our model is not completely realistic.
That defect must be corrected in an improved version.

\section*{Acknowledgments}
A fruitful collaboration with R.C. Hwa is gratefully acknowledged. This 
work was partially supported by U.S. DOE under Grant No. 
DE-FG03-96ER40972 


\begin{figure}[t]
\psfig{figure=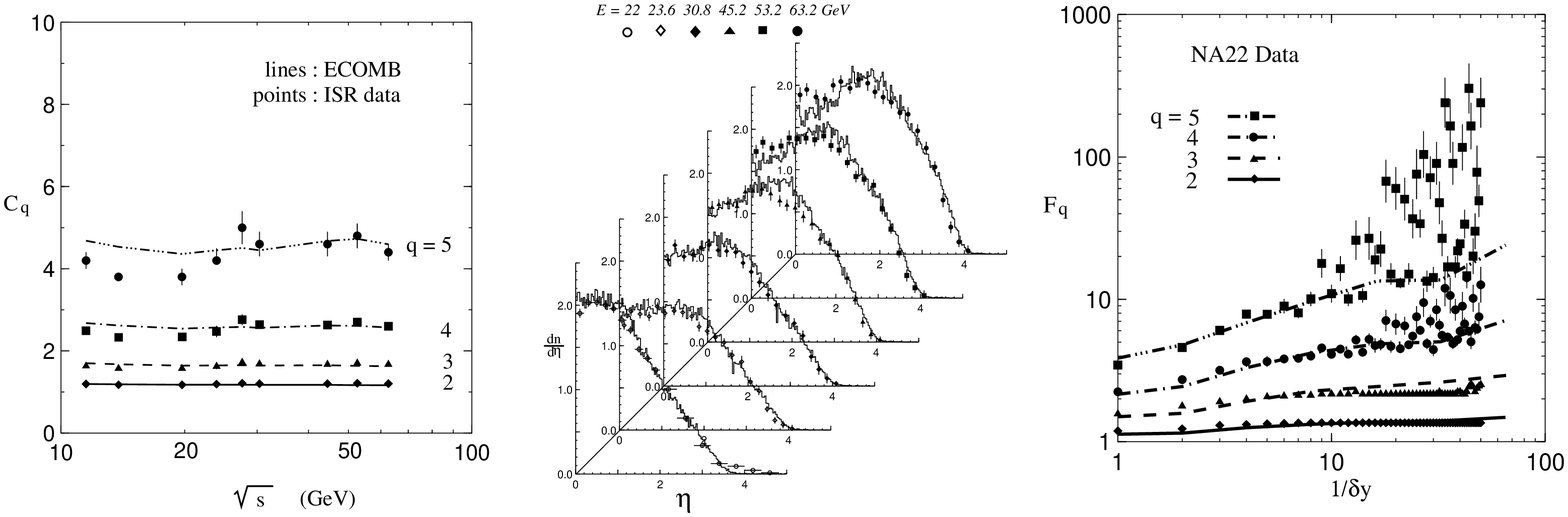,height=1.5in}
\caption{Standard moments of the multiplicity distribution. 
         The data are from Ref.\ [8]. \label{fig:n_ch}}
\end{figure}

\begin{figure}
\caption{Rapidity distributions at various cm energies.
Symbols are the data from Ref.\ [10], while the histograms are the
result of ECOMB.}
\end{figure}

\begin{figure}
\caption{Intermittency data of normalized factorial
moments for $q=2-5$. The data are from Ref.\ [11]. The lines are
determined from ECOMB.}
\end{figure}

\end{document}